\documentclass{ws-procs9x6}
\newcommand{\nwc}{\newcommand}
\nwc{\eq}{Eq.~}
\nwc{\eqs}{Eqs.~}
\nwc{\fig}{Fig.~}
\nwc{\figs}{Figs.~}
\nwc{\se}{Sec.~}
\nwc{\Tint}[1]{{\hbox{$\sum$}\!\!\!\!\!\!\int}_{\!\!\!\!#1}}
\renewcommand{\vec}[1]{{\mathbf{#1}}}
\nwc{\bfx}{{\bf x}}
\nwc{\bfi}{{\bf i}}
\nwc{\tinymsbar}{{\overline{\mbox{\tiny\rm{MS}}}}}
\nwc{\nl}  {\newline}
\nwc{\be}  {\begin{equation}}
\nwc{\ee}  {\end{equation}}
\nwc{\bmu} {\bar{\mu}}
\nwc{\ba}  {\begin{eqnarray}}
\nwc{\ea}  {\end{eqnarray}}
\nwc{\nn}  {\nonumber\\}
\nwc{\Tr}  {\mathop{\rm Tr}}
\nwc{\re}  {\mathop{\rm Re}}
\nwc{\im}  {\mathop{\rm Im}}
\nwc{\Hc}  {\mathop{\rm H.c.}}
\nwc{\la}[1]{\label{#1}}
\nwc{\rmi}[1]{{\! \mbox{\scriptsize #1}}}
\nwc{\nr}[1]{(\ref{#1})}
\nwc{\fr}[2]{{\frac{#1}{#2}}}
\nwc{\msbar}{\overline{\mbox{\rm MS}}}
\nwc{\lambdamsbar}{\Lambda_{\overline{\rm MS}}}
\nwc{\Nf}{N_{\rm f}}
\nwc{\Nc}{N_{\rm c}}
\def\slash#1{#1\!\!\!/\!\,\,} 
\def\lsi{\raise0.3ex\hbox{$<$\kern-0.75em\raise-1.1ex\hbox{$\sim$}}}
\def\gsi{\raise0.3ex\hbox{$>$\kern-0.75em\raise-1.1ex\hbox{$\sim$}}}
\nwc{\lsim}{\mathop{\lsi}}
\nwc{\gsim}{\mathop{\gsi}}
\begin{document}

\title{Mesonic Correlators in Hot QCD}

\author{M. Laine}

\address{Faculty of Physics, University of Bielefeld,\\ 
D-33501 Bielefeld, Germany \\
E-mail: laine@physik.uni-bielefeld.de}

\maketitle

\abstracts{
Certain spacelike mesonic correlation lengths serve as 
interesting theoretical probes for the reliability of perturbation 
theory in high-temperature QCD, are directly sensitive 
to chiral symmetry restoration and to the axial anomaly, and might
also have indirect phenomenological signatures in idealised heavy ion 
collision experiments. I review here
the weak-coupling predictions for some of these correlation lengths, 
to be compared with results from lattice Monte Carlo simulations.}

\section{Introduction}

At a finite temperature $T$, Lorentz symmetry is broken 
by the heat bath, so that temporal and spatial directions 
are, in general, unrelated.
Correlation functions in the (Minkowski) time direction can be used to define, 
after a Fourier transform, spectral functions for various operators, 
and the spectral functions in turn determine fundamental ``real-time''
properties of the plasma, such as particle production rates.\cite{rates} 
These are then
directly measurable in heavy ion collision
experiments (modulo problems with background contamination, etc). 

Correlation functions 
in spatial directions, in contrast, address questions 
such as: At which length scales are thermal fluctuations correlated? 
At which length scales are external charges screened? In principle 
these ``static'' observables are also physical and may lead to detectable
signals, however in practice the relations are indirect and therefore 
weaker than for real-time observables. On the other
hand, static quantities are eminently suited to measurements 
in {\em lattice experiments}. Lattice simulations will hence be the 
most immediate ``phenomenological'' reference point in the following.

There is naturally a vast variety of different operators that can 
be correlated. The operators can be classified according to their discrete
and continuous global symmetry properties,
leading to many independent correlation lengths. The analytic
structures of various purely {\em gluonic} correlators were 
discussed in Ref.~\refcite{ay}, and the corresponding 
lattice measurements
are rather precise by now.\cite{adjoint}\cdash\cite{mu}
The purpose of this talk is to analyse the structures 
of {\em mesonic} correlators, for which the status has been 
somewhat less advanced. 
This talk is based on the 
original study in Ref.~\refcite{lv}.

%
\section{Detailed setup}

As just mentioned, 
we will focus on correlation 
lengths $\xi$ of mesonic observables, that is operators 
of the type $\mathcal{O} =  \bar\psi\, \Gamma F^{a} \psi$, where 
\ba
 \Gamma & = & \{1,\gamma_5,\gamma_\mu,\gamma_\mu\gamma_5 \}  
 \;, \\
 F^{a} & = & 
 \{ F^s, F^{n} \}, \; 
 F^s \equiv {\mathbb I}_{\Nf\times \Nf}, \;
 F^n = {\rm traceless} ~\Nf\times \Nf~ {\rm matrix}
 \;, 
\ea
$\Nf$ is the number of quark flavours, 
and all colour, spinor and flavour indices have been suppressed.
We may recall that at least some among these operators have direct 
physical significance, for instance
\ba
  \bar\psi \gamma_5 F^{s} \psi & \propto & \eta'\mbox{-meson}
 \;, \\  
  \bar\psi \gamma_5 F^{n} \psi & \propto & \mbox{pion} 
 \;, \\
  \bar\psi \gamma_0 F^{s} \psi & \propto & \mbox{baryon number density}
 \;, \\
  \bar\psi \gamma_0 F^{n} \psi & \propto & \mbox{electric charge density} 
                                         ~~(\mbox{for}~\Nf = 3)
 \;.
\ea
The lattice determination of spatial correlation
functions for operators of this type was pioneered by 
DeTar and Kogut\cite{dtk} and others\cite{mtc} a long time ago,
but quantitatively significant results relevant for the physical
infinite volume chiral continuum limit, are only being 
produced presently.\cite{pdf}\cdash\cite{review}
In order to streamline the discussion, we will 
restrict to one of the operators in 
the following, 
namely the interpolating field for 
the pion: $\pi \sim i \bar\psi \gamma_5 F^n \psi$.
For $\bar\psi \gamma_0 F^{s} \psi$, see \se\ref{se:fs}, 
and for the other operators, Ref.~\refcite{lv}. 

The spatial correlation function
of the pion field has at large distances ($|\vec{x}|\to\infty$)
the structure  
\ba
 C_\vec{x} \equiv 
 \int_0^{1/T} \! {\rm d}\tau \, 
 \langle \pi(\tau,\vec{x}) \pi(0,\vec{0}) \rangle
 \sim \frac{1}{|{\bf x}|^n} \exp\left( - \frac{|{\bf x}|}{\xi(T)}\right) 
 \;,
 \la{space} \la{Cx}
\ea
where $\tau$ is the Euclidean time coordinate of the imaginary
time formulation. 
In \eq\nr{space},
$n$ is some unspecified power, 
while $\xi(T)$ defines the correlation length
we are interested in. We will denote the
inverse of the correlation length by
$m \equiv \xi^{-1}$, and call it a ``screening mass''.

The question now is, how does $m(T)$ behave as the temperature 
is increased from below to above the critical temperature $T_c$ of the 
chiral phase transition? At small temperatures, chiral symmetry is broken, 
and the pion is massless in the chiral limit, 
resulting in an infinite correlation length, or $m=0$. 
At high temperatures, on the other hand, the screening mass
approaches $2\pi T$, as we will recall presently. Thus the pion
screening mass provides for a finite and gauge-invariant 
``order parameter'' for chiral symmetry restoration. 

The reason that the screening mass equals $2\pi T$ at high
temperatures, is easily understood. For $T \gg T_c$, asymptotic freedom
sets in, and the correlator can be determined in perturbation theory. 
A computation of the leading order graph, \nl
%
\begin{minipage}[t]{15cm}

\centerline{\epsfysize=2.0cm%
\epsfbox[100 650 300 750]{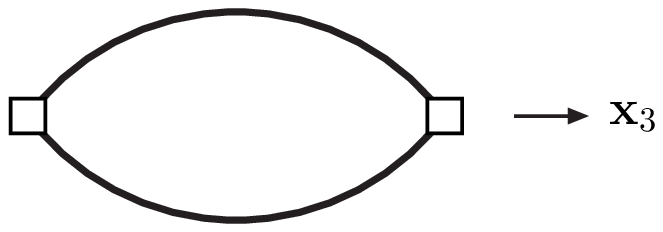}~~\hspace*{2cm}}

\end{minipage}
%
and a subsequent Fourier transform
[$C_\vec{p} = \int_{\tau,\vec{x}}
\exp(-i\omega_n\tau-i\vec{p}\cdot\vec{x})C_\vec{x}$], yield
\be
 C_\vec{p} \sim 
 \Tr [ F^{n} F^{n} ] \, \Nc \,
 T \!\! \sum_{n=-\infty}^{\infty} 
 \frac{i |\vec{p}|}{8\pi} \ln\frac{2 \omega_n - i |\vec{p}|}
 {2 \omega_n + i |\vec{p}|} \;, \; \;\;
 \omega_n  = 2 \pi T(n + \fr12) 
 \;.
\ee
Now, the singularity closest to the origin in momentum space
(determining the behaviour at the largest distances
in configuration space) 
is seen to be 
a two-particle threshold at $|\vec{p}| = \pm 2 i \omega_0$.
Referring for a moment to the spatial directions in a 
(2+1)-dimensional language, with $x_3$ as the ``time'' coordinate, 
both quarks are on-shell 
at this point and have a minimal ``energy'' $ip_3 = \pm \omega_0$ 
[the minimal energy appears
for a momentum $\vec{p}_\perp = \vec{0}$, where 
we denoted $\vec{p}\equiv (\vec{p}_\perp,p_3)$]. Thus, taking
an inverse Fourier transform back to configuration space, 
we obtain the exponential decay advertised above, 
$C_\vec{x} \sim \exp(-2\omega_0 |\vec{x}|)$.

%
\section{Next-to-leading order for flavour non-singlet correlators}

The question we would like to address next is, what is the 
first weak-coupling correction to $m = 2\pi T$? This
computation has been described
in Ref.~\refcite{lv}, and we only reiterate the main steps here.  

For flavour non-singlet operators, the relevant graphs might 
appear to be  \nl
%
\begin{minipage}[t]{15cm}

\vspace*{0mm}

\centerline{\epsfysize=2.0cm%
\epsfbox[50 650 400 750]{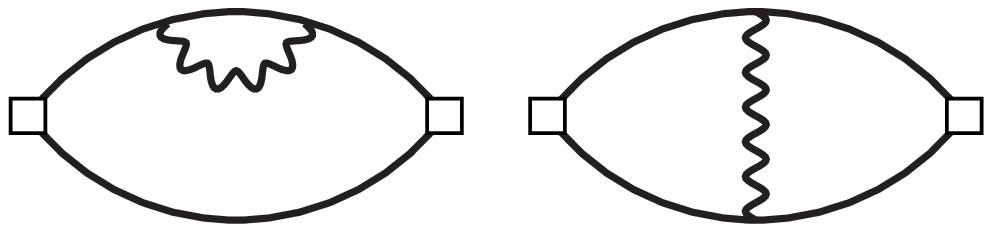}~~\hspace*{3cm}}

\vspace*{-2mm}

\end{minipage}
%
but in fact there are infinitely many higher order graphs that need to be 
taken into account, symbolically of the types \nl
%
\begin{minipage}[t]{15cm}

\vspace*{0mm}

\centerline{\epsfysize=2.0cm%
\epsfbox[100 650 260 750]{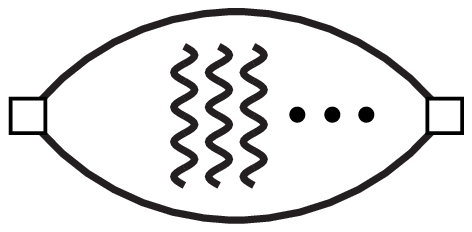}~~\hspace*{2cm}}

\vspace*{-2mm}

\end{minipage}
%
These are not suppressed, because near the two-particle threshold,
the quarks are almost on-shell,  
$p^2 = \omega_n^2 + \vec{p}_\perp^2 + p_3^2  \approx 0$,
and consequently the dimensionless expansion parameter
associated with adding a further line can be seen to be
$\mathcal{O}(g^2T/|\slash{p}|)\sim \mathcal{O}(1)$.
In other words, the situation is analogous to the 
computation of the energy of a two-quark bound state at zero temperature, 
where again infinitely many graphs contribute. 

The analogy with the bound-state system suggests 
immediately a tool for organising the computation: 
As in the study of quarkonia at $T=0$, energies of bound
states can be addressed with
an effective theory called ``Non-Relativistic QCD'' (NRQCD).\cite{nrqcd} 
NRQCD techniques were first employed in the present context 
by Huang and Lissia.\cite{hl} 

In order for an effective description to apply, the system
must possess a scale hierarchy. Let us recall why one exists here, at 
high enough temperatures where the QCD coupling constant $g$ is small. 
The basic point is that because quarks of a definite Matsubara
frequency $\omega_n$ interact with bosonic Matsubara zero-modes only, 
we expect that their off-shellness is related to the momentum scales of 
the latter: $|i p_3\pm  \omega_n| \lsim gT$. More precisely, we may
recall that the static potential has in (2+1) dimensions the structure
\be
 V(\vec{x}_\perp) \sim g^2 T \ln |\vec{x}_\perp|
 \;.
\ee
Thus the typical (transverse) momentum $\vec{p}_\perp$ of the bound-state
constituents satisfies 
\be
 \vec{p}_\perp^2/\omega_n \sim V(\vec{x}_\perp)
 \Rightarrow \vec{p}_\perp^2 \sim (gT)^2 
 \;,
\ee
and the binding energy is (dropping again possible logarithms)
\be
 |i p_3\pm  \omega_n| \sim \vec{p}_\perp^2 / \omega_n \sim V(\vec{x}_\perp) 
 \sim g^2T
 \;.
\ee
We thus indeed find a scale hierarchy 
\be
 |i p_3 \pm \omega_n| \ll |\vec{p}_\perp| \ll \omega_n
 \;.
\ee 

The effective action for this kinematic range 
can most easily be constructed by choosing a convenient basis 
for Dirac matrices (making $\gamma_0\gamma_3$ diagonal), 
and rewriting then the Dirac spinor as
\be
  \psi \equiv \left( 
  \begin{array}{l} \chi \\ \phi \end{array}
  \right)
 \;, 
\ee
where $\chi$, $\phi$ are two-component spinors.
In this basis the pion operator becomes 
$ \pi \sim i (\chi^\dagger \sigma_3 \phi - \phi^\dagger \sigma_3 \chi)$,
where $\sigma_3$ is a Pauli matrix. Restricting to the Matsubara 
mode $\omega_0$ and expanding the QCD action
to order $1/\omega_0$, 
the (on-shell) effective Lagrangian reads 
\be
 \mathcal{L} = 
 i \chi^\dagger \biggl(
 \omega_0 - g A_0 + D_3 
 - \frac{\nabla_\perp^2}{2 \omega_0} 
 \biggr) \chi + 
 i \phi^\dagger \biggl(
 \omega_0 - g A_0 - D_3 
 - \frac{\nabla_\perp^2}{2 \omega_0} 
 \biggr) \phi
 \;, \la{Leff}
\ee 
where $D_3$ is a covariant derivative.
We may note that:
\begin{itemize}

\item 
Both $A_0$ and $A_3$ play an important dynamical role. 

\item
The transverse gauge fields $A_1,A_2$, 
on the other hand, can be ignored, as long as we are
interested in an energy shift of $\mathcal{O}(g^2T)$:
they are of higher order than $\nabla_\perp \sim gT$. 
(They would be of order unity with respect to the magnetic 
scales $\nabla_\perp \sim g^2T$, but that gives an energy correction
of order $\mathcal{O}(g^4T)$ only.)

\item 
To be consistent at $\mathcal{O}(g^2T)$, we should 
replace $\omega_0$ of the tree-level effective Lagrangian
by a matching coefficient
$M = \omega_0 + \mathcal{O}(g^2T)$ 
that needs to be determined. 

\end{itemize}

While the final 
value of the matching coefficient $M$ is unambiguous, there
are many ways to determine it. In order to avoid computing wave 
function normalisation factors, we determine $M$ by matching
on-shell quark self-energies in the original QCD as well as 
in the effective theory of \eq\nr{Leff}. 
On the side of QCD, a simple 1-loop graph, \nl
%
\begin{minipage}[t]{15cm}

\centerline{\epsfysize=2.0cm%
\epsfbox[100 680 250 750]{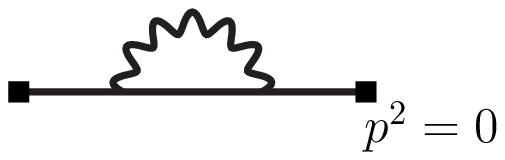}~~\hspace*{2cm}}

\end{minipage}
%
produces near the pole the inverse propagator 
\be
 \Sigma(p) = i \slash{p} - i g^2 C_F \left.
 \Tint{q} \frac{\gamma_\mu (\slash{p} 
 - \slash{q})_f\gamma_\mu}
 {(p-q)^2_f (q^2 + \lambda^2)_b} \right|_{p^2=0}  
 \;, \la{prop}
\ee
where $C_F = (\Nc^2 - 1)/ 2 \Nc$, $\Nc$ is the number of colours,
$\Sigma\!\!\!\!\!\int\,$ is the standard imaginary-time integration measure, 
$(...)_f$ and $(...)_b$ denote fermionic and bosonic Matsubara
four-momenta, respectively, and we have introduced
a gluon mass $\lambda$ as an intermediate infrared (IR) regulator.
It can be seen that the integral
in \eq\nr{prop} remains finite for $\lambda \to 0$, and then, 
up to order $\mathcal{O}(g^2)$, 
solving for the zero 
of \eq\nr{prop} is equivalent 
to solving 
\be
 p^2 + m_\rmi{eff}^2 = 0 
 \;, \la{meff}
\ee 
where $m_\rmi{eff}^2 = g^2T^2 C_F/4$
is nothing but the ``hard effective mass'' of quarks
introduced recently in Ref.~\refcite{amy}.

On the side of NRQCD, the computation of the quark self-energy
is to be carried out
order by order in $1/\omega_0$.\cite{am} It is easy to see that, 
in fact, 1-loop corrections then vanish,\footnote{%
  It is important to 
  note that at this point we are using un-resummed gluon 
  propagators both on the QCD and on the effective theory side.
  } 
such that the on-shell
point is determined directly by the mass scale $M$
appearing in the tree-level propagator, $ip_3 = \pm M$. 

Consequently, solving for the zeros in $ip_3$ of \eq\nr{meff}
to order $\mathcal{O}(g^2T)$,
we see that $M$ needs to match $\omega_0 + m_\rmi{eff}^2/2 \omega_0$,
leading to
\be
  M = \omega_0 + g^2 T \frac{C_F}{8\pi} + \mathcal{O}(g^4 T)
  \;. \la{eq:M}
\ee
This value replaces $\omega_0$ in the first terms inside 
the parentheses in \eq\nr{Leff}.

The parameters of the effective theory having been determined, 
it remains to solve its dynamics to order $g^2T$. We define 
the correlation function
\be
 C(\vec{r},x_3)\equiv
 \int_{\vec{R}} \, 
 \left\langle
 \phi^\dagger\left(\vec{R}+\frac{\vec{r}}{2},x_3\right) \sigma_3
 \chi\left(\vec{R}-\frac{\vec{r}}{2},x_3\right) 
 \;
 \chi^\dagger (0) \sigma_3
 \phi(0) 
 \right\rangle 
 \;, \la{Cr} 
\ee
and integrate out $A_0,A_3$.
To order $\mathcal{O}(g^2)$, $C(\vec{r},x_3)$ satisfies
the partial differential equation
\be
 \Bigl[
 \partial_{x_3} + 2 M - \frac{1}{\omega_0} \nabla_\vec{r}^2 + V(\vec{r})
 \Bigr] C (\vec{r},x_3) \propto \delta(x_3)\delta(\vec{r})
 \;, \la{Schr}
\ee
where
\ba 
 V(\vec{r}) & = & 
 g^2T C_F \int \frac{{\rm d}^{2-2\epsilon} q}
 {(2\pi)^{2-2\epsilon}} 
 \left\{
 \frac{ 1 - e^{i \vec{q}\cdot\vec{r}}}{q^2 + \lambda^2} 
 - \frac{ 1 + e^{i \vec{q}\cdot\vec{r}}}{q^2 + m_{D}^2} 
 \right\}  
 \la{Vr}
 \\ 
 & = & 
 g^2T \frac{C_F}{2\pi}
 \left[
 \ln\frac{m_{D} r}{2} + \gamma_E - K_0 (m_{D} r) 
 \right] \;.
 \la{Vr2}
\ea 
On the last line we took the continuum limit
(${\epsilon\to 0}$) and removed the IR regulator from 
the propagator of $A_3$ (${\lambda\to 0}$). Moreover, 
$K_0$ is a modified Bessel function, and $m_{D}$
is the Debye mass appearing in the propagator of $A_0$, 
$m_D^2 = g^2T^2(\Nc/3 + \Nf/6)$.
Solving for the lowest energy level
of the Schr\"odinger equation 
following from \eq\nr{Schr} at $x_3 \neq 0$, 
we finally obtain 
\ba
 m & = & 2 \pi T + g^2T \frac{C_F}{2\pi} 
 \Bigl( \fr12 + \hat E_0 \Bigr) \;, \quad
 \hat E_0^{\Nf = 0} =   0.164 
  \;,  \la{eq:mfinal}
\ea
where $\hat E_0$ depends mildly on $\Nf$ (cf.\ Ref.~\refcite{lv}), and 
we have shown explicitly the value relevant for the quenched 
theory.\footnote{%
  To be precise, the quenched theory value applies in the
  ``perturbative'' broken Z($\Nc$) vacuum, i.e. the one where the 
  phase of the Polyakov loop is trivial. In the unquenched theory
  this vacuum constitutes in any case the global ground state of the system.}

Note that 
in \eq\nr{eq:mfinal}, the factor $1/2$ inside the parentheses comes
from the ``constituent mass'' correction in~\eq\nr{eq:M}, while 
$\hat E_0$ comes from the solution of the Schr\"odinger equation. 
If the solution corresponds to a bound state, 
why then is $\hat E_0$ positive? The answer lies simply in the form 
of the potential in~\eq\nr{Vr2}: it is a logarithmically ``confining''
potential, with $V(\vec{r}\to\vec{0})\to -\infty$, 
$V(\vec{r}\to\vec{\infty}) \to +\infty$, 
whose overall normalisation just happens to be such that 
the lowest energy eigenvalue is positive. 

\begin{figure}[tb]


\centerline{
~~~\epsfysize=5.0cm\epsfbox{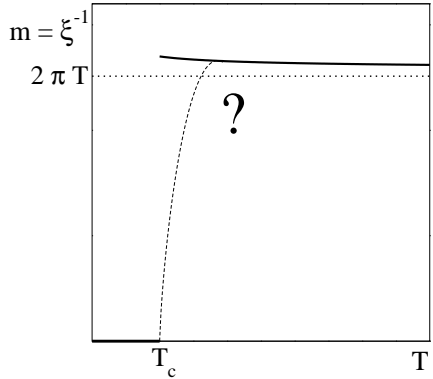} }


\caption{The pion screening mass
as a function of the temperature. The dashed line is a guess for
how the low-$T$ and high-$T$ limits might be 
connected, and in the continuous form drawn assumes a second
order transition (i.e.\ $\Nf = 2$). Note, however, that no sign of 
the indicated overshooting of $2\pi T$ has been observed in lattice 
simulations.}

\la{fig:mfinal}
\end{figure}

The result of \eq\nr{eq:mfinal} is plotted, 
with a certain choice of renormalisation scale for $g^2$
(cf.\ Ref.~\refcite{adjoint}), in \fig\ref{fig:mfinal}. 
We note that the correction appears small even at realistic $T$, 
but positive in sign. The change of $m$ 
across $T_c$ is thus very rapid --- it is not smoothed
by the $\mathcal{O}(g^2)$ term like for the pressure,\cite{es}
but rather made more pronounced. Close enough to $T_c$ the mass
of course has to decrease, either continuously
(the case for $\Nf = 2$, indicated with a dashed 
line in \fig\ref{fig:mfinal}), 
or with a small discontinuity (for $\Nf = 0, 3$).

For the pressure, the next-to-next-to-leading order correction, 
$\mathcal{O}(g^3)$, comes with an opposite sign to $\mathcal{O}(g^2)$,
and appears very large.\cite{jk} (At the same time the ultimate 
result, including higher orders still,\cite{pressure} resembles more  
the $\mathcal{O}(g^2)$ approximation.) 
It would be interesting to find out 
whether there is an $\mathcal{O}(g^3)$
correction in the pion correlation length as well. 

%
\section{Flavour-singlet correlators}
\la{se:fs}

For flavour singlets the pattern is very different from 
the flavour non-singlets, as the singlet operators couple to 
purely gluonic ``glueball'' states. For baryon density, for instance, 
one finds that\cite{mu,db} 
\ba
 \bar\psi \gamma_0 F^s \psi 
 & \longleftrightarrow & 
 - \frac{i\Nf }{3\pi^2} g^3 \Tr[ A_0^3 ]
 \;, \la{nB} 
\ea
or graphically, that the correlation function is dominated
by graphs like \nl
%
\begin{minipage}[t]{15cm}

\vspace*{-1mm}

\centerline{\epsfysize=2.0cm%
\epsfbox[100 650 250 750]{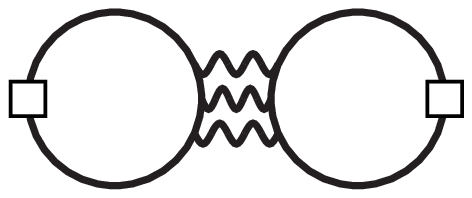}~~\hspace*{2.5cm}}

\vspace*{-2mm}

\end{minipage}
%
The corresponding screening
mass has been determined some time ago,\cite{mu}
and evaluates to $m(\Tr[A_0^3])\approx 5T$ at $T \approx 2 T_c$.
The inverse, $\xi \approx 1/5T$, is then the distance scale at 
which fluctuations in baryon number density are correlated. 
The glueball operators mediating the correlations of a number of 
other mesonic flavour-singlets have been worked out 
in Ref.~\refcite{lv}.

%
\section{On U$_A$(1) symmetry at high temperatures}

The U$_A$(1) axial symmetry is broken explicitly by the 
anomaly. It has been suggested,\cite{ua1} however, that it might get 
``effectively restored'' somewhat above the critical temperature, 
say at $T \gsim 2 T_c$. What is meant by this is that the 
topological susceptibility, $\chi = \langle\nu^2\rangle/V$, where 
$\nu$ is the topological charge of the gauge configuration, or 
the net number of zero-modes of the Dirac operator,  
and $V$ is the volume, decreases 
rapidly at temperatures above $T_c$. This becomes quite obvious
particularly  
if the number of colours is increased above $\Nc = 3$; 
see, e.g., Ref.~\refcite{dd} and references therein. Consequently, 
it could be expected that anomalous effects cease to operate. 

It may be appropriate to remark at this point, though, that it is 
possible to find observables for which the
U$_A$(1) symmetry appears not to get restored. Consider, for instance, 
the screening masses for the operators 
$V_3^s = \bar\psi \gamma_3 \psi$, 
$A_3^s = \bar\psi \gamma_3\gamma_5 \psi$,
measured from correlation functions in the $x_3$-direction.
Now, U$_A$(1) symmetry restoration would imply that the 
correlation functions for $V_3^s,A_3^s$ are identical and, 
therefore, that the screening masses are identical. Yet, 
a perturbative analysis can be used to indicate\cite{lv} that
$A_3^s$ couples to the glueball-like state 
$g^2 \epsilon_{ij3}\Tr[ A_0F_{ij}]$ of the 
dimensionally reduced theory,\cite{dr}
and has thus a non-vanishing screening mass 
$m = m_{D} + \mathcal{O}(g^2T)$,\cite{ay,adjoint}
where the next-to-leading order correction is also 
known numerically.\cite{hp} 
At the same time, the correlator related to $V_3^s$ is 
exactly conserved, and has no non-trivial screening mass
associated with it. Therefore, the  U$_A$(1) symmetry 
appears broken even at high temperatures.

%
\section{Conclusions}

Various mesonic correlation lengths are well-defined 
gauge-invariant physical observables that seem to be computable, 
for flavour non-singlets, at least
up to next-to-leading order in the gauge coupling $g^2$, 
as outlined in this talk. 
Higher orders could also be reached in principle. For flavour 
singlets, on the other hand, the correlation lengths can be related
to those of purely gluonic states, which have been determined 
previously from the dimensionally reduced theory with good 
numerical precision.\cite{mt,hp,mu}

The comparison of these analytic predictions with results from 
four-dimensional lattice simulations poses an interesting puzzle. 
Indeed, lattice results so far are consistent with the leading order 
value $m = 2 \pi T$, as soon as the temperature is a bit above
the critical one,\cite{pdf}\cdash\cite{review} 
while the next-to-leading order correction 
advertised above is positive, requiring an ``overshooting''
of $m=2\pi T$. Lattice results involve systematic
errors, though, so that the discrepancy cannot be considered too
serious at present.

On the side of analytic efforts, one 
could envisage a number of computations which might
facilitate the comparison with ever more precise future simulations.  
For instance, the leading discretization effects 
could be determined analytically, even though this of course
is specific to the fermion action used. (Different groups have
rather different preferences here.) Another point of 
possible relevance is that some of the correlation functions
involve an ${x_3}$-dependent prefactor in front of the exponential, 
and it might be useful to take it into account in the fitting procedure.

To conclude, let us recall that the 
ultimate theoretical goal of the exercise discussed here is 
to estimate the reliability of perturbation theory at finite $T$, 
in order to then use perturbation theory 
with more confidence for other observables 
for which lattice is not well suited. Given that good QCD predictions 
can eventually be obtained for static correlation lengths and that 
there is a remarkably stark difference between the two phases, 
one is however also lead to wonder, once again, whether 
some interesting phenomenological signatures might be found
for these quantities.

%
%
%
%
%
%
%
\section*{Acknowledgments}

I wish to thank M.~Veps\"al\"ainen for collaboration on the work
presented in this talk, and S.~Datta, E.~Laermann, G.D.~Moore and
R.D.~Pisarski for useful comments related to topics covered in it. 


\end{document}